\documentclass[runningheads]{llncs}
\usepackage{tabularray}
\usepackage{amsfonts}
\usepackage{amssymb}
\usepackage[T1]{fontenc}
\usepackage{multirow}
\usepackage{amsmath}
\usepackage{marvosym}
\usepackage{stackengine}
\usepackage{algorithm}
\usepackage{algpseudocode}
\usepackage{subcaption}
\usepackage{caption}
\usepackage{xcolor}
\usepackage[colorlinks, citecolor=cyan]{hyperref}
\usepackage{graphicx}

\begin{document}
\title{Improved Esophageal Varices Assessment from Non-Contrast CT Scans}
\titlerunning{Improved Esophageal Varices Assessment on Non-Contrast CT}
%
%
\author{Chunli Li\inst{1,2,3}$^\star$\and Xiaoming Zhang\inst{2,3}$^{(\textrm{\Letter},\star)}$\and Yuan Gao\inst{2,3}$^\star$\and Xiaoli Yin\inst{1}\and Le Lu\inst{2}\and \\ Ling Zhang\inst{2} \and  Ke Yan\inst{2,3}$^\textrm{\Letter}$ \and Yu Shi\inst{1}$^\textrm{\Letter}$}


%
\authorrunning{C. Li, X. Zhang, Y. Gao, et al.} 
\institute{Department of Radiology, Shengjing Hospital of China Medical University,
110004, Shenyang, China \and DAMO Academy, Alibaba Group \and
Hupan Lab, 310023, Hangzhou, China \\
\email{zxiaoming360@gmail.com}; \email{yanke.yan@alibaba-inc.com}; \email{18940259980@163.com}
}
%
%
\maketitle             
\renewcommand{\thefootnote}{\fnsymbol{footnote}}
\footnotetext[1]{Equal contributions. \textsuperscript{\Letter}Corresponding authors. The work was done during C. Li's internship at Alibaba DAMO Academy. }
\begin{abstract}
Esophageal varices (EV), a serious health concern resulting from portal hypertension, are traditionally diagnosed through invasive endoscopic procedures. Despite non-contrast computed tomography (NC-CT) imaging being a less expensive and non-invasive imaging modality, it has yet to gain full acceptance as a primary clinical diagnostic tool for EV evaluation. To overcome existing diagnostic challenges, we present the Multi-Organ-cOhesion-Network (MOON), a novel framework enhancing the analysis of critical organ features in NC-CT scans for effective assessment of EV. Drawing inspiration from the thorough assessment practices of radiologists, MOON establishes a cohesive multi-organ analysis model that unifies the imaging features of the related organs of EV, namely esophagus, liver, and spleen. This integration significantly increases the diagnostic accuracy for EV. We have compiled an extensive NC-CT dataset of 1,255 patients diagnosed with EV, spanning three grades of severity. Each case is corroborated by endoscopic diagnostic results. The efficacy of MOON has been substantiated through a validation process involving multi-fold cross-validation on 1,010 cases and an independent test on 245 cases, exhibiting superior diagnostic performance compared to methods focusing solely on the esophagus (for classifying severe grade: AUC of 0.864 versus 0.803, and for moderate to severe grades: AUC of 0.832 versus 0.793). 
To our knowledge, MOON is the first work to incorporate a synchronized multi-organ NC-CT analysis for EV assessment, providing a more acceptable and minimally invasive alternative for patients compared to traditional endoscopy. 

\end{abstract}

\section{Introduction}

Esophageal varices (EV) are a significant complication often stemming from chronic liver diseases, with portal hypertension being the pivotal cause driving blood into smaller collaterals within the lower esophagus~\cite{garcia2017portal}, potentially leading to life-threatening hemorrhages and shock. Detecting EV at an early stage is difficult, with confirmation usually requiring invasive endoscopic procedures. Given that the extent of venous dilation correlates with hemorrhage risk, a precise assessment of EV severity is vital.

Invasive endoscopy is the conventional method for EV evaluation and, despite its effectiveness, it is associated with discomfort, potential complications like infection and bleeding, and higher healthcare costs. As a less invasive approach, dynamic contrast-enhanced CT (DCE-CT) allows for detailed visualization of EV and related collaterals, improving patient compliance and reducing discomfort and bleeding risks linked to endoscopy. However, methods such as Yan et al.'s~\cite{yan2022novel}, which use radiomic features from DCE-CT, have limited generalizability, and iodinated contrast agents pose risks of adverse reactions.

Non-contrast CT (NC-CT) offers a quick and low-radiation alternative, chest NC-CT scanning can be valuable for grading EV. Liver characteristics such as fibrosis and alterations in volume can indicate the pressure within the portal venous~\cite{shi2021three}, which is critical in appraising EV risk and severity. Portal hypertension-induced changes in the spleen, such as enlargement, indirectly reflect this pressure and suggest EV bleeding risk. Therefore, a combined analysis of the liver, spleen, and esophagus could improve the understanding and grading of EV risks, and enhance clinical decisions. However, it faces several hurdles in EV assessment: differentiation difficulties due to lower contrast resolution, limited visibility of smaller varices, and the inconsistent distribution of varices throughout the esophagus.

Previous attempts to evaluate EV have focused on DCE-CT imaging of the liver, spleen, esophagus, and portal venous system's vascular features~\cite{luo2023clinical,wan2022quantitative,yan2022novel}. However, no current grading methods fully consider the esophagus's close relationships with adjacent structures. To optimize the assessment of EV, we have drawn upon radiologists' comprehensive diagnostic techniques and physicians' clinical diagnostic knowledge, introducing the framework of Multi-Organ-cOhesion-Network (MOON) based on chest NC-CT. Our approach enhances the accuracy of EV evaluation and surpasses previous radiomics methods reliant on DCE-CT imaging.

MOON adopts a holistic strategy, simultaneously analyzing the regions of interest (ROIs) from the NC-CT scans of these organs. It employs a multi-organ framework that incorporates the Organ Representation Interaction (ORI) module, allowing features from the liver and spleen to inform the analysis of esophageal characteristics at each step within the network. Additionally, the Hierarchical Feature Enhancement (HFE) module is designed to highlight the esophagus' distinct morphology by synthesizing features across various levels, focusing on areas that exhibit abnormal varices. To achieve uniformity in decision-making across the various organ domains, which are vital for an accurate final diagnosis,  we incorporate ordinal regression loss~\cite{cheng2008neural} and canonical correlation analysis (CCA)~\cite{andrew2013deep} loss into our training process. This synergistic use of loss functions fosters a cohesive feature integration, thereby reducing inconsistencies in the decision-making process and enhancing the overall diagnostic performance.

Our studied dataset is sufficiently extensive, comprising chest NC-CT scans from 1255 patients diagnosed with EV, each confirmed via endoscopic examination and categorized into three differentiated levels of severity. The goal of MOON is to predict the grade of EV. It exhibits strong diagnostic proficiency, attaining an AUC performance of \textbf{0.864} versus 0.736 using DCE-CT~\cite{yan2022novel} for severe EV cases and, \textbf{0.832} versus 0.802 using DCE-CT~\cite{wan2022quantitative} for moderate to severe cases, showcasing its effectiveness in evaluating EV from NC-CT imagery.

\section{Method}

\begin{figure}[!ht]
    \centering
    
    \begin{subfigure}[b]{1\linewidth}
        \centering
        \includegraphics[width=1\linewidth]{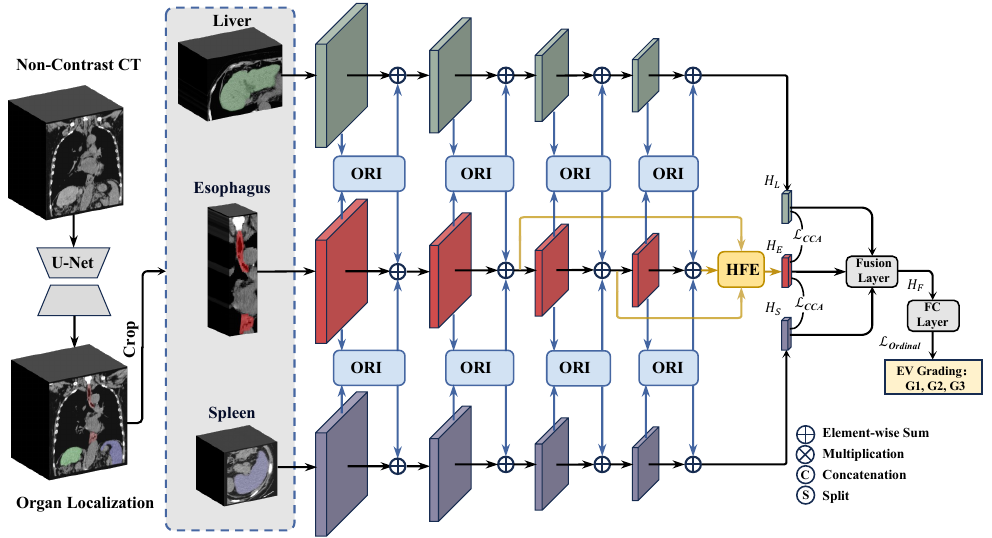}
        \caption{}
        \label{fig:MOON}
    \end{subfigure}
    \begin{subfigure}[b]{1\linewidth}
        \centering
        \includegraphics[width=1\linewidth]{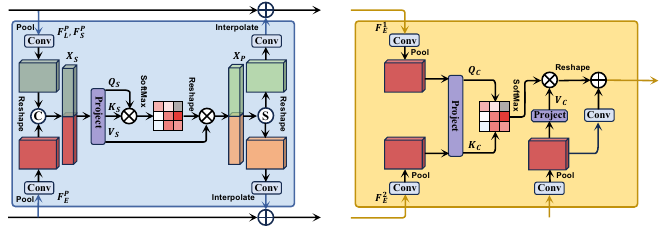}
        \captionsetup{font=scriptsize,labelformat=empty}
        \caption[]{(b) \hspace*{20em} (c)}
        \label{fig:module}
    \end{subfigure}
    
    \caption{An overview of Multi-Organ-cOhesion-Network (MOON). (a) The pipeline of multi-stage pipeline. (b) Organ representation interaction (ORI). (c) Hierarchical feature enhancement (HFE).}
    \label{pipeline}
\end{figure}

Deep learning, particularly convolutional neural networks (CNNs)~\cite{he2016deep,lecun1989handwritten,lecun1989backpropagation}, have significantly advanced the detection of pathologies or diseases with subtle imaging differences in Hounsfield Units on NC-CT images, often imperceptible to radiologists~\cite{wasserthal2022totalsegmentator}. There are four key designs in our method: (1) To assess EV, a global receptive field is essential. Uniformer~\cite{li2023uniformer} combines convolution with self-attention, which is utilized to adeptly capture esophageal features, providing the comprehensive coverage necessary for accurate EV evaluation. (2) An effective EV assessment necessitates combining data from multiple organs. Our ORI module facilitates this, integrating liver and spleen features with those of the esophagus to improve EV detection. (3) The diversity in varices sizes and esophageal anatomy requires a multi-scale strategy. We answer this need with the HFE module, which assimilates features at various scales to enhance classification adaptability. (4) For effective integration across multiple organ branches and precise identification of the severity grades of EV, we train our model using a hybrid loss function that combines ordinal regression and CCA. This approach enables the model to maintain the ordinal relationships inherent to the various EV severity levels while enhancing the correlation between multi-organ features, reducing bias from reliance on any single organ, and improving the overall diagnostic accuracy and generalizability of the model.

\subsection{Multi-Organ-cOhesion-Network (MOON)}

\textbf{Pipeline of MOON.} The UniFormer architecture~\cite{li2023uniformer}, combining local and global feature extraction, addresses the medical imaging challenge of accurately assessing EV by providing a comprehensive receptive field suitable for the esophagus's unique structure, thereby enabling precise identification of esophageal features without the constraints of traditional 3D CNNs~\cite{isensee2021nnu,tran2018closer} or Vision Transformers~\cite{dosovitskiy2020image}. MOON, depicted in Fig.~\ref{pipeline}\textcolor{red}{(a)}, is a multi-stage, multi-organ framework incorporating UniFormer as its core for each branch. It starts with an off-the-shelf nnUNet~\cite{isensee2021nnu} for organ segmentation to localize the esophagus, liver, and spleen. Post-segmentation, respective image data are fed dedicated branches to extract features across scales. Within this framework, the ORI module plays a pivotal role, adeptly handling the unique anatomical features of the organs, e.g., the tendency for spleen enlargement and liver lesions in the portal venous system. Its capacity to engage with each organ's distinct properties is vital for accurate EV assessment. The HFE module, in turn, is specifically designed to enhance the esophagus branch,  which is essential given that experiments in Sec.~\ref{exp} indicate esophageal features are most indicative of EV. Through its focus on the esophagus, the HFE module promotes effective feature integration and sharpens the identification of key esophageal details for precise EV detection. 

\noindent\textbf{Organ Representation Interaction.}
To address the nuanced diagnosis of EV stemming from portal hypertension, a comprehensive CT scan analysis of the liver and spleen is required due to the subtle morphological and texture changes they undergo. As shown in Fig.~\ref{pipeline}\textcolor{red}{(b)}, the ORI module is designed to enhance the interpretation of these images by refining liver and spleen imaging features into a coherent form, directing attention to critical areas. It streamlines the imaging features of these organs, amplifying essential characteristics and employing a self-attention mechanism that operates on tokenized 3D representation to link esophageal features to hepatic and splenic indicators of EV. Given that the features of the esophagus, liver, and spleen at each stage, defined as $\textbf{F}_{E}\in\mathbb{R}^{H_{e}\times W_{e}\times D_{e} \times C}$, $\textbf{F}_{L}\in\mathbb{R}^{H_{l}\times W_{l}\times D_{l} \times C}$, and $\textbf{F}_{S}\in\mathbb{R}^{H_{s}\times W_{s}\times D_{s} \times C}$, respectively. $\textbf{F}_{E}^{P}, \textbf{F}_{L}^{P}, \textbf{F}_{S}^{P} \in \mathbb{R}^{H_{n}\times W_{n}\times D_{n} \times C}$ are pooled features. An attention-ready form $X_S \in \mathbb{R}^{H_{n}W_{n}\times D_{n}\times 2C}$ is produced by concatenation and reshaping. The \textit{query} ($Q_S$), \textit{key} ($K_S$), and \textit{value} ($V_S$) matrices are generated from $X_S$ through learned transformations. The scaled dot-product attention is applied as $\text{Attention}(Q_S, K_S, V_S) = \text{SoftMax}\left(\frac{Q_{S}K_{S}^T}{\sqrt{d_k}}\right)V_S$, with $d_k$ representing the dimensionality of the keys, allowing the model to capture feature interactions across the global scope of the input space. The outputs from the attention mechanism are aggregated into a tensor $T_S$ and then projected to an enhanced feature space, resulting in $X_P = T_{S}W_P$, where the projection matrix $W_P$ is of size $\mathbb{R}^{2C\times 2C}$. This process consolidates the learned information of adjacent organs and infuses them into the esophageal representation, thereby strengthening the subtle interconnections specific to EV. By finally splitting and then reintegrating $X_P$, followed by refinement through convolution and subsequent interpolation, ORI module achieves an interwoven feature representation.

\noindent\textbf{Hierarchical Feature Enhancement.}
The HFE module refines the central branch that analyzes esophageal features, sharpening the detection of localized characteristics indicative of EV. It advances beyond former multi-scale fusion methods~\cite{lin2017feature,tan2020efficientdet} by implementing a cross-attention mechanism. This feature promotes interaction among different depth levels and conveys detail-rich information from the shallow layers, imperative for EV detection, to the in-depth features that represent localized details within an expansive anatomical context. Illustrated in Fig.~\ref{pipeline}\textcolor{red}{(c)}, through a process of convolution, pooling, and cross-attention, the HFE module adeptly blends shallow and deep features. This strategic combination solidifies the main branch's feature representation, bolstering the network's overall diagnostic accuracy. The HFE module uses convolution and pooling to adjust the dimensions of features from varying depths, starting with the intermediate $\textbf{F}_{E}^{1}$ and moving to the deeper $\textbf{F}_{E}^{2}$, ultimately aligning them with the dimensions of the deepest esophagus feature: $\textbf{F}_{E}^{3} \in \mathbb{R}^{H_{n}\times W_{n}\times D_{n} \times C}$. After pooling, these features are reorganized into an attention-ready form $X_C \in \mathbb{R}^{H_{n}W_{n}\times D_{n}\times C}$. Using a form of cross-attention, the attention mechanism then transforms $X_C$ to generate the \textit{query} ($Q_C$), \textit{key} ($K_C$), and \textit{value} ($V_C$) matrices, where $Q_C$ originates from the intermediate features $\textbf{F}_{E}^{1}$, $K$ from the deeper features $\textbf{F}_{E}^{2}$, and $V$ from the deepest features $\textbf{F}_{E}^{3}$. The scaled dot-product attention is applied as $\text{Attention}(Q_C, K_C, V_C) = \text{SoftMax}\left(\frac{Q_{C}K_{C}^T}{\sqrt{d_k}}\right)V_C$. The HFE module orchestrates the selective enhancement of local features and their integration with deeper, more globally aware features. By fusing shallow and deep features through the HFE module, with attention-focused features refined by point-wise convolutions for a more robust representation.

\noindent\subsection{Training Paradigm of MOON}

MOON is proposed and designed to tackle the intricacies of EV diagnosis by integrating multi-organ information. Within this scope, we deploy an ordinal regression loss, symbolized as $\mathcal{L}_{\text{Ordinal}}$, which is applied to the unified logits $\mathbf{H}_F$. These logits consolidate data from the esophagus, liver, and spleen branches, in line with~\cite{cheng2008neural}. The ordinal regression loss is crucial for measuring the model's capability to accurately classify the different stages of EV, as it verifies the alignment between $\mathbf{H}_F$ and the ground truth labels $\mathbf{Y}$. Additionally, the framework integrates CCA loss $\mathcal{L}_{\text{CCA}}$, from multi-view learning, which is key to maximizing the correlation between two feature sets for improving cross-modal retrieval~\cite{andrew2013deep}. Unlike approaches that seek complementary features, MOON employs $\mathcal{L}_{\text{CCA}}$ to capitalize on the interconnectedness of organ features by amplifying their correlation. Specifically, it targets the alignment of logits from the esophageal branch ($\mathbf{H}_E$) with those from the liver ($\mathbf{H}_L$) and spleen ($\mathbf{H}_S$) branches. By serving as a regularization mechanism, $\mathcal{L}_{\text{CCA}}$ facilitates the convergence of characteristics derived from multiple organs, thus enhancing the model's capability to interpret interrelated organ information, as shown in Algorithm~\ref{cca}.
\begin{algorithm}
\caption{Canonical Correlation Analysis Loss}
\begin{algorithmic}[ht!]
\Statex \textbf{Input}: Two relevant logits of $H_{1}, H_{2}\in\mathbb{R}^{n,h}$, $\epsilon=10^{-12}$. 
\Statex \textbf{Output}: CCA loss $\mathcal{L}_{CCA}(H{1}, H_{2})$  optimizing correlation between $H_{1}$ and $H_{2}$.
\State $H_{1}, H_{2} \gets \frac{H_{1} - \text{mean}(H_{1})}{\text{std}(H_{1}) + \epsilon}, \frac{H_{2} - \text{mean}(H_{2})}{\text{std}(H_{2}) + \epsilon}$ 
\State $C_{1}, C_{2} \gets \frac{H_{1}^{T} H_{1}}{n-1}, \frac{H_{2}^{T} H_{2}}{n-1}$ \Comment{Compute covariance}
\State $\lambda_1, V_1 \gets \text{eig}(C_{1}); \lambda_2, V_2 \gets \text{eig}(C_{2})$ \Comment{Eigen-decomposition}
\State $H_{1}, H_{2} \gets H_{1} V_1[:, \text{top } h], H_{2} V_2[:, \text{top } h]$ \Comment{Project to top eigenvectors}
\State $C \gets \frac{1}{n-1} H_{1}^T H_{2}$ \Comment{Compute cross-covariance}
\State $\mathcal{L}_{CCA}(H_{1}, H_{2}) \gets -\frac{\text{Tr}(C)}{\|H_{1}\|_F \cdot \|H_{2}\|_F + \epsilon}$ \Comment{Calculate canonical correlation}
\State \textbf{return} $\mathcal{L}_{CCA}(H_{1}, H_{2})$
\end{algorithmic}
\label{cca}
\end{algorithm}

The total loss $\mathcal{L}_{\text{Overall}}$ for a given batch is a composite of the $\mathcal{L}_{\text{Ordinal}}$ and $\mathcal{L}_{\text{CCA}}$, weighted appropriately, that not only accounts for accurate staging of EV but also reinforces the interconnectedness of the multi-organ representations:
\begin{equation}
\mathcal{L}_{\text{Overall}} = \lambda \cdot \mathcal{L}_{\text{Ordinal}}(\mathbf{H}_F, \mathbf{Y}) + (1-\lambda) \cdot (\mathcal{L}_{\text{CCA}}(\mathbf{H}_E, \mathbf{H}_L) + \mathcal{L}_{\text{CCA}}(\mathbf{H}_E, \mathbf{H}_S)).
\label{lcca}
\end{equation}

\section{Experiments}\label{exp}

\textbf{Datasets.} 
We collected a dataset in a retrospective manner, with NC-CT images stratified into three grades of EV severity as outlined by~\cite{dahong2000trial,kawano2008short}: $G1$ for \textit{Mild} ($F1$, $RC\text{-}$), $G2$ for \textit{Moderate} ($F1$, $RC+$ or $F2$, $RC\text{-}$), $G3$ for \textit{Severe} ($F2$, $RC+$, $F3$, $RC+$ or $F3$, $RC\text{-}$). This dataset comprises images from 1,010 patients diagnosed with varying stages of EV ($G1$: 331, $G2$: 252, $G3$: 427), acquired with Philips Ingenuity 4 and Siemens Sensation 64 CT scanners. All patients were confirmed to have EV through endoscopy, and all CT scans were performed within one month prior to the endoscopic examination. The collected images were processed using a standardized abdominal window setting. To rigorously evaluate the proposed method, a 5-fold cross-validation approach was employed. Additionally, an independent test cohort comprising 245 subjects was curated from a single center to substantiate the method's validity. Eligibility for data inclusion required being over 18 years old and not having received treatment before the CT scans.

\noindent\textbf{Implementation Details.} 
Throughout the training process, several data augmentation techniques were applied, such as random rescaling, flipping, and cutout operations. For localizing the esophagus, liver, and spleen, we utilized the pre-trained nnUNet~\cite{isensee2021nnu} configured for low-resolution image processing as our localization network. Once localized, the ROIs for the esophagus, liver, and spleen were resized to dimensions of $40\times40\times100$, $256\times196\times36$, and $152\times196\times24$, respectively. To construct each branch of the MOON, we employed the Uniformer-S model~\cite{li2023uniformer}, which was pre-trained on the Kinetics-600~\cite{carreira2018short}. The MOON framework was trained employing the Adam~\cite{kingma2014adam} optimizer with an initial learning rate set to $10^{-5}$. The training spanned across 100 epochs, with the learning rate being reduced by a factor of two every 20 epochs, following a piece-wise constant decay schedule. We set $\lambda=0.8$ for Eqn.~(\ref{lcca}). The framework was implemented using Pytorch and run on Nvidia V100 GPUs.

\noindent\textbf{Main Experimental Results.}
Table~\ref{tab:main} consolidates experimental results demonstrating that for evaluating EV, single features derived from the esophagus alone surpass those from the liver or spleen in terms of accuracy. This underscores the critical importance of the esophagus in EV detection. As a result of portal hypertension, there is a marked enlargement of the spleen, potentially leading to changes in the internal textural characteristics of the organ. These alterations can be detected by neural networks. Hence, in terms of accuracy during 5-fold cross-validation, spleen features rank second, marginally surpassing those from the liver. Compared to single-organ input, integrating data from multiple organs achieved a better outcome with an increase of 0.003 to 0.039 in AUC for the evaluation of G3 on an independent test set. However, the performance improvement for assessing $\geq$G2 was not significant. This may be due to the fact that solely relying on post-fusion strategies does not adequately and effectively integrate features originating from multiple organs. The incorporation of the ORI and HFE modules into MOON has led to a performance that outperforms simple post-fusion strategies. In an independent test set, the AUC for G3 evaluation saw an increase of 0.006-0.040, and for the single-organ esophageal model, there was a boost of 0.038-0.061. Moreover, in the assessment of $\geq$G2, there was also a significant enhancement in AUC (0.016-0.034) relative to simple post-fusion strategies.In the fusion layer, we employ several strategies, with concatenation~\cite{feichtenhofer2016convolutional} emerging as the most effective, particularly in the challenging G3 on the independent test set. G3 grades show higher detection accuracy compared to $\geq$G2, likely due to G3's clearer clinical features. Additionally, FiLM~\cite{perez2018film} stands out in 5-fold validation, achieving the highest accuracy. These findings emphasize the significance of advanced post-fusion techniques in enhancing the accuracy of multi-organ EV, further augmented by ORI and HFE.
\begin{table*}[ht!]
\caption{Comparative analysis between single-organ and multi-organ methodologies, examining various strategies for feature fusion. $\ddag$ denotes that without the interaction from ORI and HFE modules. ACC(\%): Accuracy Values; AUC: Area under Curve Values. Standard variation of the values are in $\%$. }
\resizebox{\textwidth}{!}{
\begin{tabular}{c|cccc|cccc}
\hline
\multirow{3}{*}{Methods} & \multicolumn{4}{c|}{5-fold ($n=1010$)}                             & \multicolumn{4}{c}{Independent test ($n=245$)}                       \\ \cline{2-9} 
                         & \multicolumn{2}{c|}{ $\geq$G2} & \multicolumn{2}{c|}{G3} & \multicolumn{2}{c|}{$\geq$G2}       & \multicolumn{2}{c}{G3} \\ \cline{2-9} 
                         & ACC      & \multicolumn{1}{c|}{AUC}     & ACC                & AUC                & ACC     & \multicolumn{1}{c|}{AUC}     & ACC             & AUC            \\ \hline
Single-organ (Eso.)    & $72.5_{\pm0.8}$  & \multicolumn{1}{c|}{$0.754_{\pm0.7}$} & $75.8_{\pm1.5}$            & $0.821_{\pm1.9}$            & $74.3$ & \multicolumn{1}{c|}{$0.793$} & $74.3$         & $0.803$        \\
Single-organ (Liver)        & $68.9_{\pm0.5}$  & \multicolumn{1}{c|}{$0.677_{\pm1.8}$} & $63.5_{\pm0.4}$          & $0.672_{\pm0.7}$            & $71.8$ & \multicolumn{1}{c|}{$0.679$} & $64.5$         & $0.680$        \\
Single-organ (Spleen)       & $71.0_{\pm1.0}$  & \multicolumn{1}{c|}{$0.681_{\pm1.5}$} & $61.0_{\pm2.0}$            & $0.660_{\pm1.7}$           & $68.6$ & \multicolumn{1}{c|}{$0.668$} & $64.1$         & $0.684$        \\ \hline

MOON$\ddag$ (Concat~\cite{feichtenhofer2016convolutional})     & $73.2_{\pm2.4}$  & \multicolumn{1}{c|}{$0.793_{\pm2.7}$} & $74.9_{\pm2.1}$           & $0.836_{\pm3.5}$            & 75.1 & \multicolumn{1}{c|}{0.799} & 73.9         & 0.827        \\
MOON$\ddag$ (PredSum~\cite{simonyan2014two})    & $75.4_{\pm3.2}$  & \multicolumn{1}{c|}{$0.776_{\pm3.3}$} & $76.2_{\pm3.6}$            & $0.830_{\pm3.6}$            & 75.5 & \multicolumn{1}{c|}{0.805} & 71.4         & 0.823        \\
MOON$\ddag$ (LowRank~\cite{liu2018efficient})    & $73.5_{\pm3.5}$  & \multicolumn{1}{c|}{$0.771_{\pm1.57}$} & $76.1_{\pm3.8}$            & $0.824_{\pm4.4}$            & 73.9 & \multicolumn{1}{c|}{0.779} & 71.4         & 0.806        \\
MOON$\ddag$ (FiLM~\cite{perez2018film})       & $74.9_{\pm1.4}$& \multicolumn{1}{c|}{$0.780_{\pm2.6}$} & $77.0_{\pm2.1}$            & $0.843_{\pm3.1}$            & 75.5 & \multicolumn{1}{c|}{0.799} & 73.9         & 0.842        \\ \hline
MOON (Concat~\cite{feichtenhofer2016convolutional})    & $75.5_{\pm2.2}$ & \multicolumn{1}{c|}{$\underline{0.820}_{\pm2.1}$} & $76.6_{\pm4.8}$            & $0.852_{\pm3.1}$            & \underline{78.0} & \multicolumn{1}{c|}{\textbf{0.832}} & \textbf{77.6}         & \textbf{0.864}        \\
MOON (PredSum~\cite{simonyan2014two})   & $75.6_{\pm2.8}$   & \multicolumn{1}{c|}{$\textbf{0.824}_{\pm2.6}$ } & $\underline{78.3}_{\pm2.6}$             & $0.851_{\pm3.1}$            & 76.0 & \multicolumn{1}{c|}{0.810} & 74.3         & 0.841        \\
MOON (LowRank~\cite{liu2018efficient})   & $\underline{76.0}_{\pm2.8}$  & \multicolumn{1}{c|}{$0.813_{\pm3.4}$} & $\textbf{79.0}_{\pm3.7}$            & $\underline{0.858}_{\pm4.6}$            & 76.8 & \multicolumn{1}{c|}{0.821} & 76.8        & 0.846       \\
MOON (FiLM~\cite{perez2018film})      & $\textbf{77.3}_{\pm2.1}$  & \multicolumn{1}{c|}{$0.810_{\pm2.6}$} & $78.0_{\pm4.8}$            & $\textbf{0.864}_{\pm4.4}$            & \textbf{78.1} & \multicolumn{1}{c|}{\underline{0.823}} & \underline{76.7}         & \underline{0.848}        \\ \hline
\end{tabular}}
\label{tab:main}
\end{table*}

\begin{figure}[ht!]
    \centering
        \centering
        \includegraphics[width=1\linewidth]{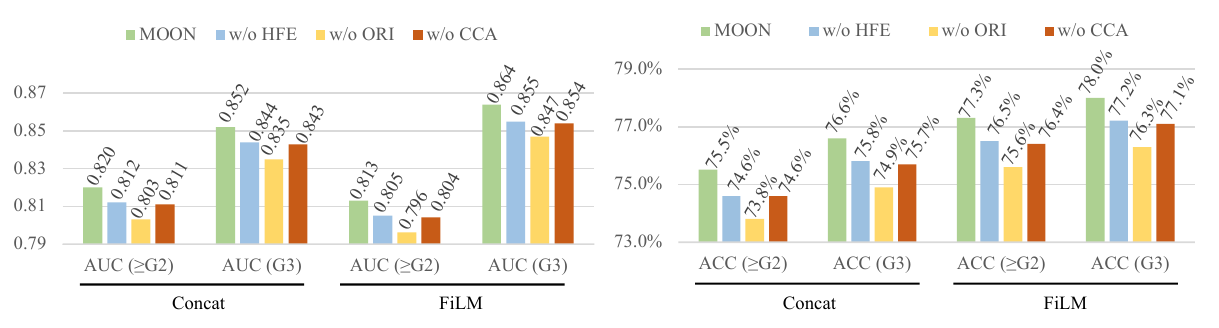}
        
        \caption{Ablation experiments on different strategies on 5-fold validation.}
        \label{fig:abla}
\end{figure}

\begin{figure}[ht!]
    \centering
        \centering
        \includegraphics[width=1\linewidth]{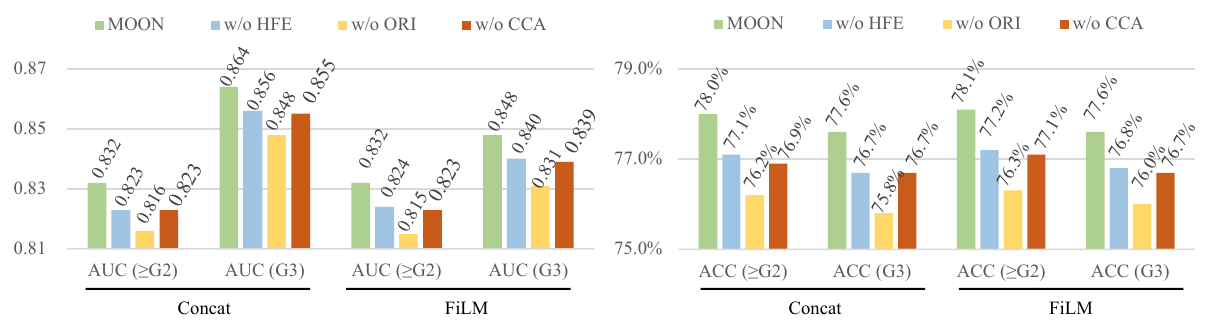}
        \caption{Ablation experiments on different strategies on independent test dataset.}
        \label{fig:abla2}
\end{figure}

\begin{figure}[ht!]
    \centering
        \centering
        \includegraphics[width=1\linewidth]{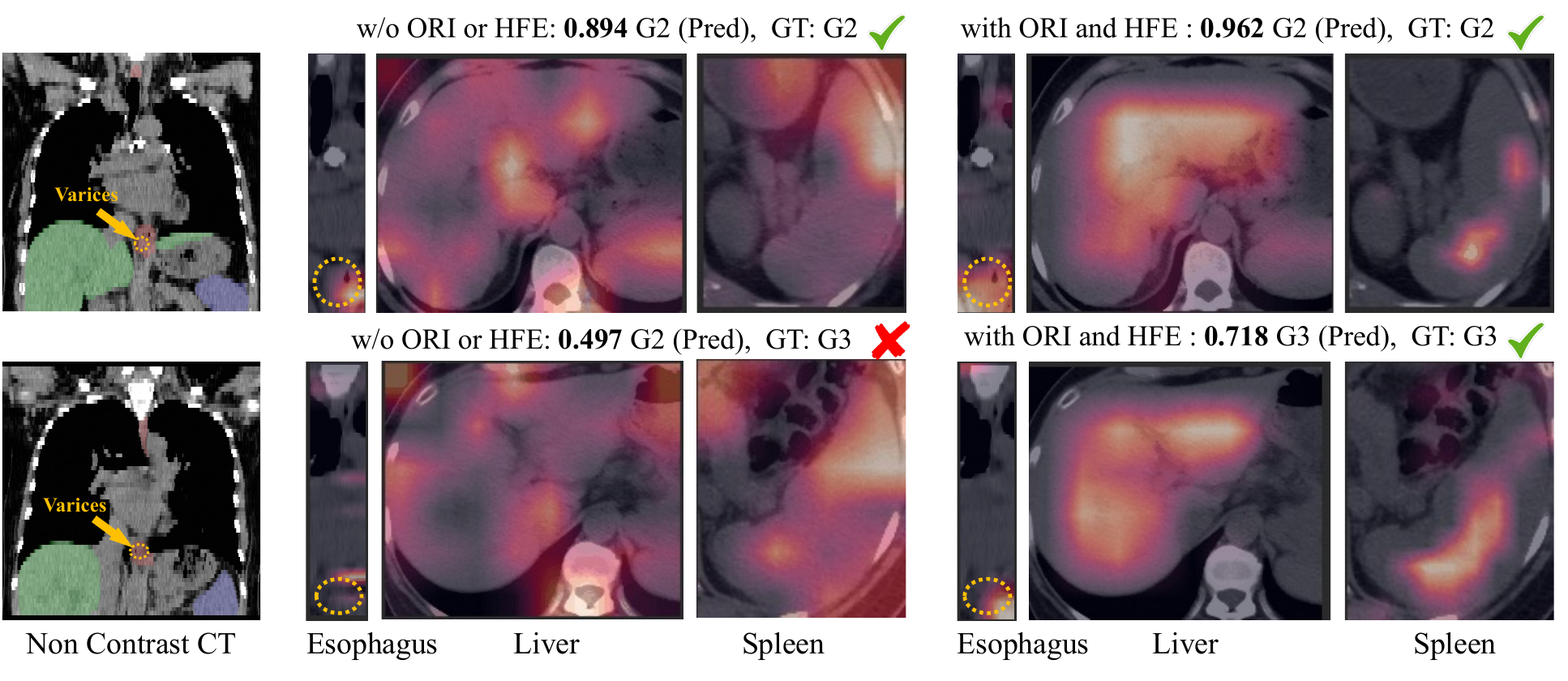}
        \caption{Comparison of Grad-CAM visualizations across different organs.}
        \label{fig:CAM}
\end{figure}

\noindent\textbf{Ablations Study.} Information regarding the comparison of EV performance evaluation during 5-fold validation with the inclusion of the ORI module, HFE module, and CCA loss strategy is provided in Fig.~\ref{fig:abla}. We discovered that the ORI module is more critical than the HFE module and CCA loss in the evaluation of EV, indicating that ORI facilitates comprehensive and effective interactive learning between the liver, spleen, and esophagus, capturing more valuable information pertinent to the assessment of EV. Additionally, other methods also hold significant value for the assessment of EV. Results of the performance on the independent test set are presented in Fig.~\ref{fig:abla2}. We observed results that were similar to those in the 5-fold validation set. Fig.~\ref{fig:CAM} highlights the role of the ORI and HFE modules in improving cross-organ interaction within the model.  Using Grad-CAM~\cite{selvaraju2017grad}, we can pinpoint the specific areas within the model's field of focus.  Correlating with endoscopic findings, we have annotated the variceal regions within the esophagus, which typically pose a detection challenge in NC-CT imaging. Without the incorporation of ORI and HFE modules for organ-specific information exchange, the model tends to overlook critical areas associated with varices, leading to potential misclassification of EV severity. In contrast, incorporating the ORI and HFE modules directs the MOON to accurately zone in on medically significant regions, and precisely targets the distended vessels in the esophagus, the hepatic portal vein, and the spleen, which are indicative of portal hypertension and EV.


\section{Conclusion}
We introduce the Multi-Organ-Cohesion-Network, designed for the non-invasive evaluation of EV using NC-CT scans. By simulating the diagnostic processes of radiologists who assess multiple organs implicated in EV, we fill the diagnostic gap in EV evaluation with NC-CT. Our extensive analysis confirms that NC-CT imaging effectively evaluates EV, outperforming previous DCE-CT-based methods. In our future work, we plan to incorporate adjunctive clinical data, e.g., blood test markers, to refine the classification efficacy of EV.

\begin{credits}
\subsubsection{\ackname}
This work was supported by the National Natural Science Foundation of China (No. 82071885); The Innovation Talent Program in Science and Technologies for Young and Middle-aged Scientists of Shenyang (RC210265); General Program of the Liaoning Provincial Education Department (LJKMZ20221160); Liaoning Provincial Science and Technology Plan Joint Foundation.
\subsubsection{\discintname}
The authors have no competing interests to declare that are relevant to the content of this article. 
\end{credits}

\bibliography{LatexSource-0915/Paper-0915}
\end{document}